\documentclass{IEEEtran}

\usepackage{cite}

\ifCLASSINFOpdf
  \usepackage[pdftex]{graphicx}
  \DeclareGraphicsExtensions{.pdf,.jpeg,.png}
\else
  \usepackage[dvips]{graphicx}
  \DeclareGraphicsExtensions{.eps}
\fi

\usepackage{amsmath}
\usepackage{upgreek}
\usepackage{array}
\usepackage{url}
\usepackage{siunitx}
\begin{document}

\title{The Mu3e Data Acquisition}

\author{
Heiko Augustin$^{1}$,
Niklaus Berger$^{2}$,
Alessandro Bravar$^{3}$,
Konrad Briggl$^{3,4}$, 
Huangshan Chen$^4$,
Simon Corrodi$^{5,a}$,
Sebastian Dittmeier$^{1}$,
Ben Gayther$^{6}$,
Lukas Gerritzen$^{5}$,
Dirk Gottschalk$^{1}$,
Ueli Hartmann$^{7}$,
Gavin Hesketh$^{6}$,
Marius K\"oppel$^{2}$,
Samer Kilani$^{6}$,
Alexandr Kozlinskiy$^{2}$,
Frank Meier Aeschbacher$^{7}$,
Martin M\"uller$^{2}$,
Yonathan Munwes$^{4}$,
Ann-Kathrin Perrevoort$^{1,b}$,
Stefan Ritt$^{7}$,
Andr\'e Sch\"oning$^{1}$,
Hans-Christian Schultz-Coulon$^{4}$,
Wei Shen$^{4}$,
Luigi Vigani$^{1}$,
Dorothea vom Bruch$^{2,c}$,
Frederik Wauters$^{2}$,
Dirk Wiedner$^{1,d}$,
 and
Tiancheng Zhong$^{4}$

\thanks{This work has been submitted to the IEEE for possible publication. 
Copyright may be transferred without notice, after which this version may no longer be accessible. 
Corresponding author: N. Berger (email: niberger@uni-mainz.de).}
\thanks{The authors would like to thank the members of the electronics workshops
at Heidelberg University, PSI and University College London for their
important contributions to the Mu3e DAQ system. 
The work of the Mainz group has been supported by the Cluster of Excellence ``Precision Physics,
Fundamental Interactions, and Structure of Matter'' (PRISMA EXC 1098 and
PRISMA+ EXC 2118/1) funded by the German Research Foundation (DFG) within the
German Excellence Strategy (Project ID 39083149); we are particularly grateful for
the expertise and infrastructure provided by the PRISMA detector laboratory. 
The Heidelberg groups acknowledge the support by 
   the German Research Foundation (DFG) funded Research Training Groups HighRR (GK
   2058) and ``Particle Physics beyond the Standard Model'' (GK 1994), 
   by the EU International Training Network PicoSec (grant no. PITN-GA-2011-289355-PicoSEC-MCNet),
   by the International Max Planck Research School for Precision Tests of Fundamental Symmetries (IMPRS-PTFS)
   and the Heinz-G\"otze-Stiftung.
   N.~Berger, A.~Kozlinskiy, A.-K.~Perrevoort, D.~vom~Bruch and F.~Wauters 
thank the DFG
for funding their work on the Mu3e experiment through the Emmy Noether programme.
A.~Sch\"oning and D.~Wiedner thank the DFG
for funding their work under grant no. SCHO 1443/2-1.
G.~Hesketh gratefully acknowledges the support of the Royal Society through
grant numbers UF140598 and RGF\textbackslash EA\textbackslash 180081.
The Swiss institutes acknowledge the funding support from
the Swiss National Science Foundation grants no.
200021\_137738, 200021\_165568, 200021\_172519, 200021\_182031 and 20020\_172706.
The Particle Physics Department (DPNC) of the University of Geneva gratefully acknowledges support from
from the Ernest Boninchi Foundation in Geneva. } 

\thanks{$^{1}$Physikalisches Institut, Ruprecht-Karls-Universit\"at Heidelberg,
Im Neuenheimer Feld 226, 69120 Heidelberg, Germany}
\thanks{$^{2}$Institut f\"ur Kernphysik and PRISMA$^{+}$ Cluster of Excellence,
          Johannes Gutenberg-Universit\"at Mainz, Johann-Joachim-Becherweg 45, 55128 Mainz, Germany}
\thanks{$^{3}$D\'epartement de Physique Nucl\'eaire et Corpusculaire,  
          Universit\'e de Gen\`eve,
          24, Quai Ernest-Ansermet, 1211 Gen\`eve 4, Switzerland}
\thanks{$^{4}$Kirchhoff-Institut f\"ur Physik, 
          Ruprecht-Karls-Universit\"at Heidelberg,
          Im Neuenheimer Feld 227, 69120 Heidelberg, Germany}          
\thanks{$^{5}$Institut f\"ur Teilchen- und Astrophysik, 
          Eidgen\"ossische Technische Hochsschule Z\"urich,
          Otto-Stern-Weg 5, 8093 Z\"urich, Switzerland}
\thanks{$^{6}$Department of Physics and Astronomy, 
                            University College London,
                            Gower Street, London WC1E 6BT, United Kingdom}
\thanks{$^{7}$Laboratory for Particle Physics, Paul Scherrer Institut,
                            Forschungsstrasse 111, 5232 Villigen, Switzerland}
\thanks{$^{a}$Now at Argonne National Laboratory, 
              9700 South Cass Avenue, Lemont, IL 60439, USA}
\thanks{$^{b}$Now at NIKHEF, Science Park 105, 1098 XG Amsterdam, Netherlands}
\thanks{$^{c}$Now at Aix Marseille Universit\'e, CNRS/IN2P3, Centre de Physique de Particules de Marseille,
        163, avenue du Luminy, Case 902, 13288 Marseille cedex 09, France}
\thanks{$^{d}$Now at Fakult\"at Physik, Technische Universit\"at Dortmund, 
          Otto-Hahn-Str. 4
          44227 Dortmund, Germany} }         

\maketitle

\begin{abstract}
The Mu3e experiment aims to find or exclude the lepton flavour violating decay 
$\mathbf{\mu^+\to e^+e^-e^+}$ with a sensitivity of one in 10$^{16}$ muon decays. The 
first phase of the experiment is currently under construction at the Paul 
Scherrer Institute (PSI, Switzerland), where beams with up to 10$^8$ muons 
per second are available. The detector will consist of an ultra-thin pixel 
tracker made from \emph{High-Voltage Monolithic Active Pixel Sensors} (HV-MAPS), 
complemented by scintillating tiles and fibres for precise timing measurements. 
The experiment produces about \SI{100}{Gbit/s} of zero-suppressed data which 
are transported to a filter farm using a network of FPGAs and fast optical 
links. On the filter farm, tracks and three-particle vertices are 
reconstructed using highly parallel algorithms running on graphics processing 
units, leading to a reduction of the data to \SI{100}{Mbyte/s} for mass storage 
and offline analysis. The paper introduces the system design and hardware
implementation of the Mu3e data acquisition and filter farm.
\end{abstract}

\section{Introduction}

\IEEEPARstart{T}{he} Mu3e experiment is designed to search 
for the lepton flavour violating decay $\mu^+\to e^+e^-e^+$; in the 
Standard Model of elementary particle physics, this process is very 
highly suppressed \cite{Hernandez-Tome:2018fbq,Blackstone:2019njl} -- an observation would 
be a clear sign of new physics beyond the Standard Model. 
The SINDRUM experiment at PSI 
performed a search for this decay in the 1980s and set a limit for 
the branching fraction $BF < 10^{-12}$ \cite{Bellgardt:1987du}. 
The Mu3e experiment aims to repeat this search with a sensitivity 
improved by four orders of magnitude \cite{RP, Mu3e:TDR}. The experiment 
will be performed in two phases, where the first phase will use an 
existing beam-line at PSI providing up to 10$^8$ muons/s whereas the 
second phase, aiming for the ultimate sensitivity, requires a new \emph{High 
intensity Muon Beam-line} (HiMB) with $10^9-10^{10}$ muons per second.

\begin{figure}[!b]
  \centering
  \includegraphics[width=0.49\textwidth]{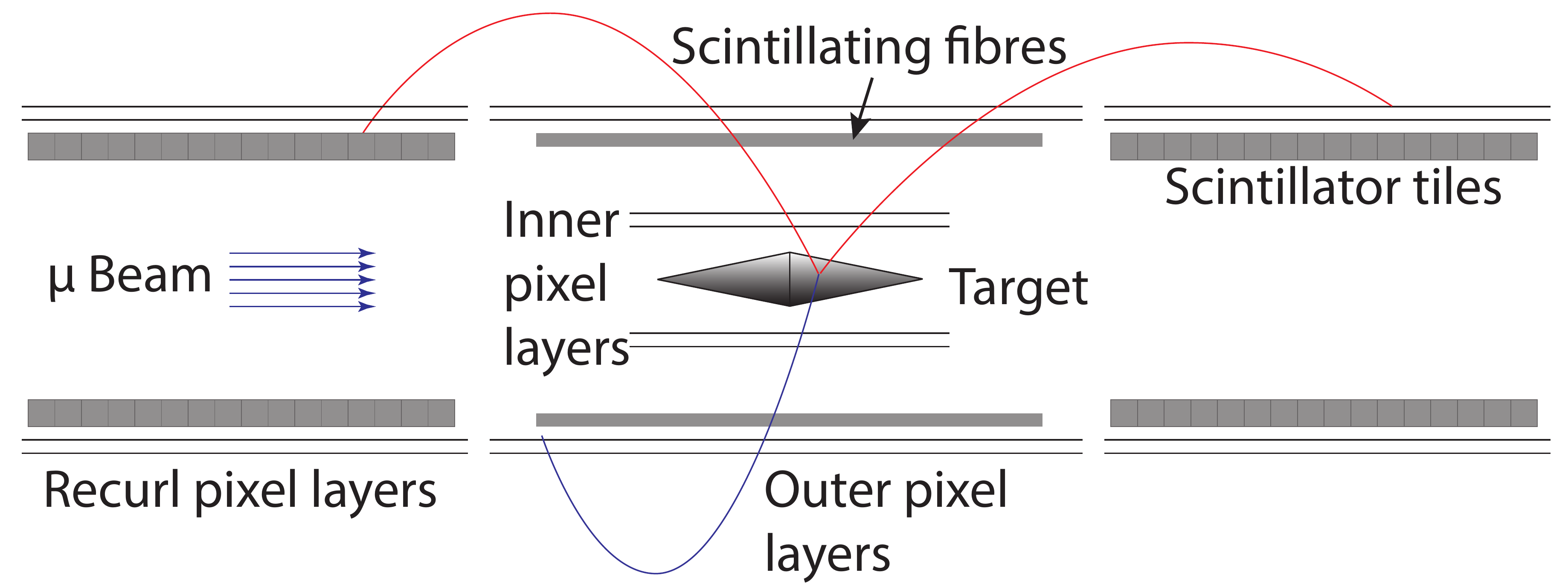}
  \caption{Schematic view of the phase~I Mu3e detector (cut along the beam 
  axis, around which it is cylindrically symmetric). Muons are stopped
  on a hollow double-cone target surrounded by two layers of vertex
  pixel sensors. Scintillating fibres provide a first timing measurement,
  followed by the outer pixel layers. In forward and backward direction,
  further pixel layers complemented by scintillating tiles greatly improve 
  the momentum and timing measurements of particles re-curling in the
  magnetic field.} 
  \label{fig:detschematic}
  \end{figure}

\begin{figure*}[!t]
  \centering
  \includegraphics[width=\textwidth]{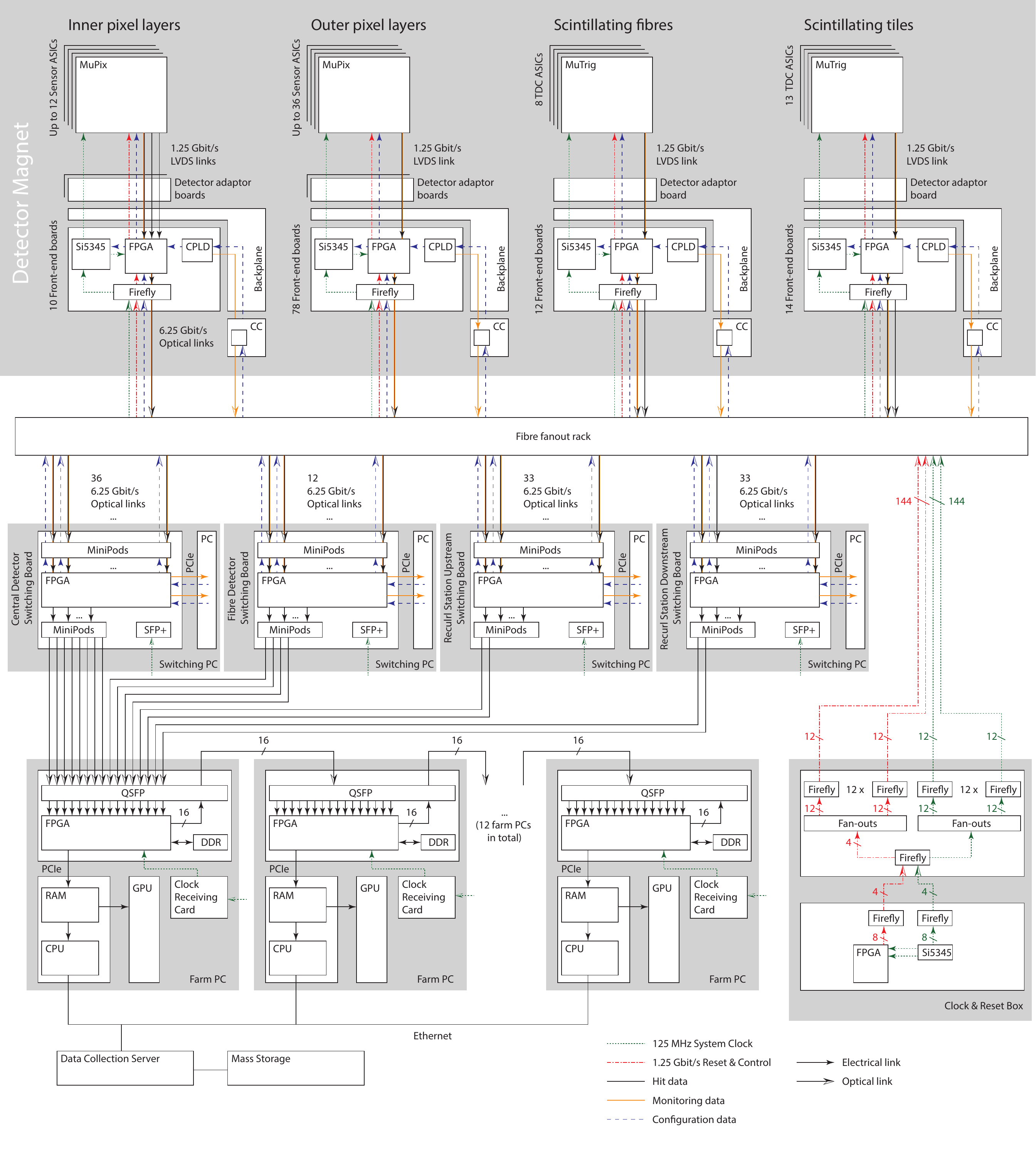}
  \caption{Overview of the Mu3e data acquisition system showing the data, clock and reset,
  configuration and monitoring connections throughout the system. Note that optical and
  electrical links are differentiated by the arrowheads.}
  \label{fig:schematic}
  \end{figure*}

The Mu3e detector has to be capable of running at these very high muon 
rates and suppressing background from both the allowed radiative decay 
with internal conversion $\mu^+\to e^+e^-e^+\nu\bar{\nu}$ \cite{Pruna:2016spf}
and accidental combinations of electrons and positrons from different 
muon decays. This requires an excellent momentum, vertex and timing 
resolution of the detector. The low momenta of the decay particles 
make multiple Coulomb scattering the dominating effect deteriorating 
the momentum resolution, which can be counteracted by minimizing the 
detector material and using an optimized tracking detector geometry.
A schematic view of the detector is shown in \figurename~\ref{fig:detschematic}. 
Extremely thin tracking layers consisting of \SI{50}{\micro\meter} 
thick HV-MAPS \cite{Peric:2007zz, Peric:2013cka,Peric:2015ska} 
mounted on polyimide-aluminium flex-prints \cite{Berger:2016lme} and cooled 
by gaseous helium are used \cite{MeierAeschbacher:2020ldo}. The HV-MAPS for Mu3e, the MuPix ASICs
\cite{Augustin:2015mqa, Augustin:2016iff, Augustin:2016hzx, Augustin:2018ppf, Augustin:2019cci, Augustin:2019qiv}, 
perform on-sensor digitization and zero-suppression and send out hit data 
via up to three \SI{1.25}{Gbit/s} \emph{Low-Voltage 
Differential Signalling} (LVDS) links. Combinatorial background 
can be further suppressed by precise timing measurements, which are 
provided by scintillating fibres (SciFi, \cite{Bravar:2017ush}) and tiles \cite{Klingenmeyer:2020gkr}, 
read out by \emph{Silicon Photomultipliers} (SiPMs). The SiPM signals are digitized 
using the custom MuTRiG ASIC \cite{Chen:2017qor, Chen:2017ria}, which also 
provides a \SI{1.25}{Gbit/s} LVDS output. The 
high granularity of the detector combined with the large particle rates produces 
an enormous amount of data; the \emph{Data Acquisition} (DAQ) system has to deal 
with roughly \SI{100}{Gbit/s} at $10^8$ muons per second. 
Table~\ref{tab:FPGABandwidthRequirements} lists the
bandwidth requirements estimated from a detailed, Geant4-based 
\cite{agostinelli:2002hh} simulation of the Mu3e detector.

\begin{table*}[t!]
  \caption{DAQ bandwidth requirements}
  \label{tab:FPGABandwidthRequirements}
  \centering
    \begin{tabular}{lrrrr}
      \hline
      Sub-detector & Maximum hit rate 	    & Maximum bandwidth needed  & Number of 			& Total expected \\
            & per front-end board   & per front-end board       & front-end boards 		& data rate	\\
            & MHz              	    & Gbit/s 					& 						& Gbit/s\\		
      \hline
      Pixels 		& 58					& 4.6 		  			    & 88					& 56 \\
      Fibres		& 28					& 2.3 		  				& 12                    & 28 \\
      Tiles		& 15					& 1.2        				& 14					& 17 \\
      \hline
      \multicolumn{5}{p{380pt}}{Estimated from the detector simulation. 
      For the fibre detector, clustering in the
      front-end FPGA is performed. For the bandwidth, \SI{75}{\percent} protocol efficiency and 
      8b/10b encoding are assumed.}
    \end{tabular}
\end{table*}

\begin{figure*}[t!]
  \centering
  \includegraphics[width=\textwidth]{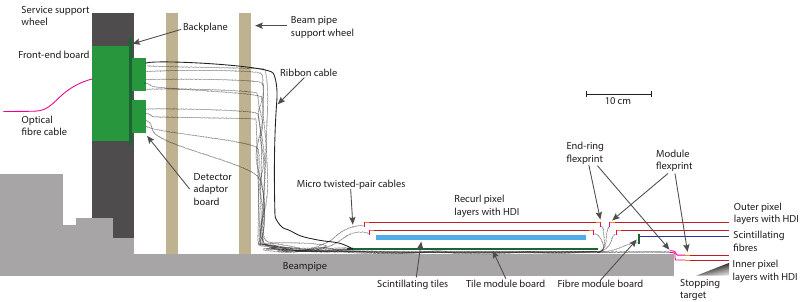}
  \caption{Cross-section of a quarter of the detector showing the active detector 
  elements and the read-out cabling -- to scale, except for the thickness of active layers,
  flexes, cables and PCBs. HDI stands for \emph{High-Density Interconnect}.
   }
  \label{fig:cabling}
  \end{figure*}

Mu3e uses a triggerless, streaming DAQ that employs a network of FPGAs and
fast optical links \cite{Bachmann:2014bfa} to transport all data out of the 
detector volume to a small farm of PCs equipped with powerful \emph{Graphics
Processing Units} (GPUs) for particle track and vertex reconstruction.
The low momentum of the decay particles combined with the strong magnetic 
field leads to strongly curved tracks. Particles can thus produce hits in 
physically distant parts of the detector; see \figurename~\ref{fig:detschematic}. 
The resulting very non-local track finding problem requires
that the data from the complete detector are available on every farm
node. The readout network thus rearranges the data such that the farm nodes
see the complete detector data of different time slices. 

The \emph{Maximum Integrated Data Acquisition System} (MIDAS, \cite{midas1997, midas:2001})
is used as the software framework for the Mu3e DAQ. All DAQ PCs run so-called
\emph{MIDAS Front-ends} interacting with the DAQ hardware either via \emph{PCIExpress} (PCIe)
or Ethernet.

The architecture of the Mu3e DAQ is shown in \figurename~\ref{fig:schematic}. In this
paper, we will discuss the path of data through the system and 
introduce the hardware used.

\section{The DAQ system}

\subsection{Front-end ASICs}

The phase~I Mu3e experiment will be built from 2844 MuPix pixel sensors complemented by 
scintillating fibres with 3072 SiPM readout channels and 5824 scintillating tiles. The pixel
sensors as well as the 278 MuTRiG ASICs reading out the scintillating detectors
send zero-suppressed hit data over 8b/10b encoded \cite{widmer1983} LVDS links.

The detector concept with a particle tracking volume outside of the detector tube
and all signal lines routed inside (together with channels for the gaseous helium cooling
system, copper bars for power and ground as well as the beam pipe) lead to very tight
space constraints for signal cabling; see \figurename~\ref{fig:cabling} for an overview. 
In the case of the pixel detector, the data first 
have to be transported out of the active tracking region, implying that material has 
to be minimized in order to reduce multiple Coulomb scattering of decay particles. The signals
are thus routed over thin aluminium-polyimide high-density interconnects \cite{MeierAeschbacher:2020ldo}
out of the active region. Flexible PCBs connected by interposers are then 
used for transferring them to micro twisted-pair cables leading to the \emph{Service Support Wheels}
(SSWs), located close to the ends of the bore of the \SI{3}{m} long \SI{1}{T} solenoid magnet.
For the inner pixel layers, three \SI{1.25}{Gbit/s} links per pixel sensor, each capable
of transmitting about \SI{30}{Mhits/s} are connected, whereas in the outer pixel layers,
only one link is connected.

For the fibre detector, one 128 channel SiPM array is connected to one SciFi module board
with four MuTRiG ASICs. These module boards are then connected via micro twisted-pair cables
to the SSW. In case of the tile detector, one MuTRiG chip reads 32 individual SiPMs and
thirteen MuTRiGs are collected on one tile module board, which is then connected to the SSW 
using a ribbon cable.

\subsection{Front-end board}

\begin{figure}
  \centering
  \includegraphics[width=\columnwidth]{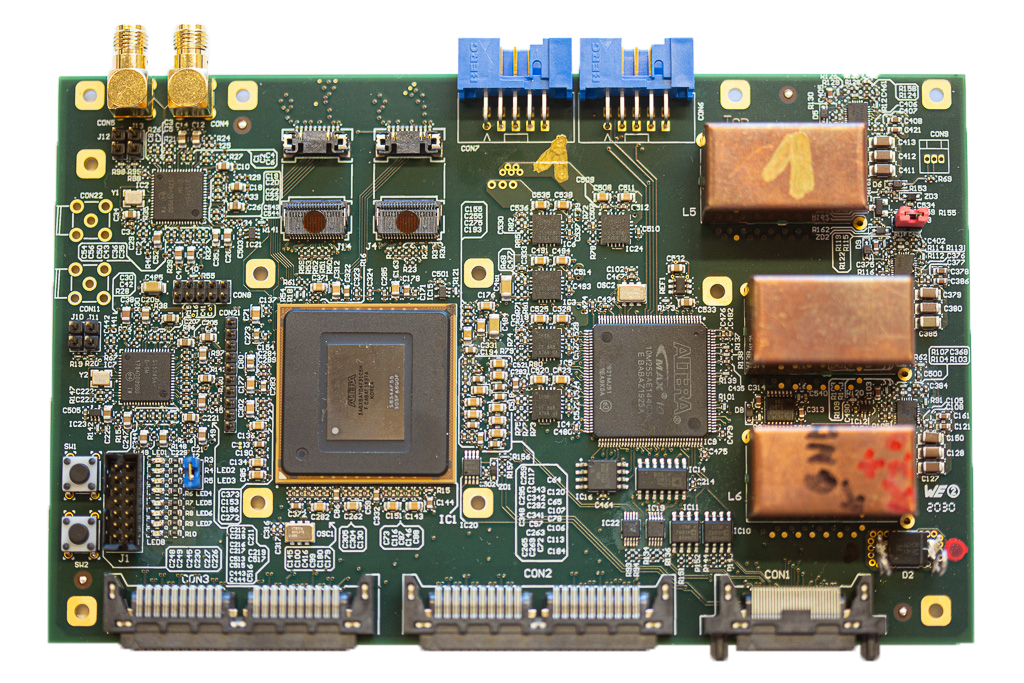}
  \caption{Front-end board. Core components are two SI5345 clock jitter cleaner ICs (top left),
  an Intel Arria~V A7 FPGA (centre left), two Samtec Firefly ECUO-B04 optical transceivers
  (connectors above the Arria~V), an Intel MAX~10 flash-based FPGA (centre right), three 
  DC-DC converters for power (dominated by the copper shielding boxes for the air coils, 
  right), JTAG connectors (top) and connectors to the backplane, both for control and 
  signals from and to the detector ASICs (bottom).
   }
  \label{fig:frontendboard}
  \end{figure}

  \begin{figure}
    \centering
    \includegraphics[width=\columnwidth]{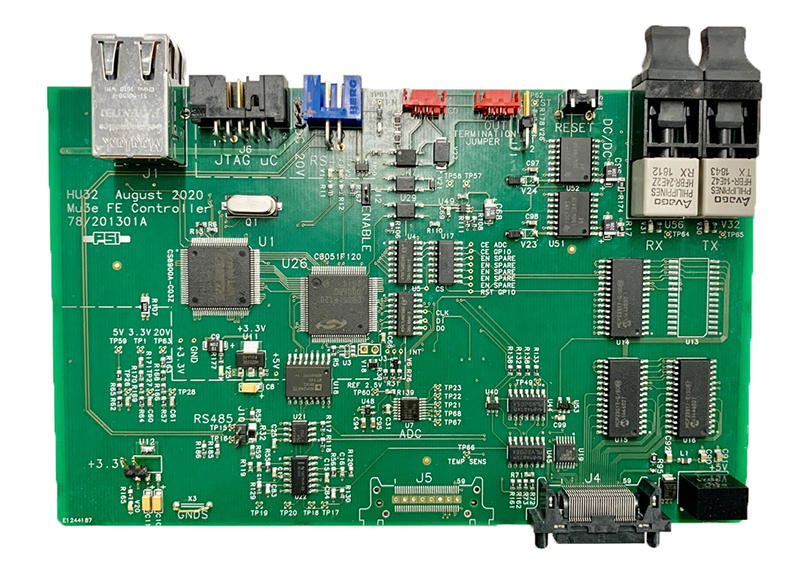}
    \caption{Crate controller card. Center left is the SiLabs C8051F120 micro-controller,
    top left the Ethernet connection, top right the optical connection and bottom right
    the connection to the backplane.
     }
    \label{fig:cratecontroller}
    \end{figure}  

A total of 114 \emph{Front-End Boards} (see \figurename~\ref{fig:frontendboard}) collect 
the sensor data, perform data processing on an FPGA and then send the data out of the 
magnet volume using \SI{6.25}{Gbit/s} optical links.
We decided to use a common front-end board for all sub-detectors and do the detector
specific signal routing on a detector adaptor board. The euro-card-sized boards sit in 
quarter-circular crates on the SSW. A backplane links up to 16 front-end boards to a crate
controller card and connects up to two detector adaptor boards to a front-end board.
The crate controller card (see \figurename~\ref{fig:cratecontroller}) uses a 
SiLabs C8051F120 micro-controller running a bare-metal C control program. 
On one side it connects to each front-end board via the backplane, and on the other side it 
connects to the experiment’s control network using the \emph{MIDAS Slow Control Bus} 
(MSCB, \cite{mscb:2001}) via either a standard Ethernet network (for tests outside the 
magnet) or an optical link. Boards can individually be power cycled in case of problems 
or firmware updates. A dedicated stand-by power allows temperature measurements in 
all front-end boards even if their power is off, which might be useful if problems with the
cooling system occur.

Data processing on the front-end board is performed by an Intel Arria~V A7 FPGA: 
The incoming data are 8b/10b decoded, and hits are separated from monitoring
information. The links are continuously monitored by detecting 8b/10b encoding and
parity errors as well as deviations from the data protocol.

For the pixel detector, a time-walk correction based on the measured time over threshold
is applied to the hit time stamps. The hits are then time sorted using insertion into memory at 
addresses determined by the hit time stamp \cite{perrevoort:2018okj}. 
A list of the hits from each ASIC and each time stamp is generated.
These lists are then processed into a single memory read sequence.
Executing this read sequence produces a single, fully time-sorted output stream containing
the data from up to 36 input links.

Due to the way the fibre ribbons are matched to the SiPM arrays, particles passing the
detector almost always trigger more than one SiPM channel. The hits of the fibre detector
are therefore clustered in time and space on the FPGA. Suppression of single hit
clusters allows to reduce the bandwidth consumed by dark counts considerably, which would 
otherwise dominate over particle signals, especially after irradiation of the SiPMs.

The resulting data streams are interleaved
with monitoring information and sent off-board using a Samtec Firefly ECUO-B04 optical
transceiver.

The Firefly transceiver offers four inputs and four outputs; one of the inputs is used
to receive the \SI{125}{MHz} system clock. Two Silicon Labs Si5345 jitter cleaners and
clock multipliers receive the clock and produce five clocks for the detector ASICs and
eleven clocks for the FPGAs. A second input is used for resets and run state transitions;
here we use a \SI{1.25}{Gbit/s}, 8b/10b encoded data stream called \emph{reset stream}, 
where the \SI{8}{bit} datagrams are encoding different transitions. 
Special care has to be taken in the firmware
to ensure that resets occur on a specific clock edge all across the system \cite{mueller2019}.
A third input is used for control and configuration information, e.g.~the threshold tune
values of all the connected pixels. This link runs at \SI{6.25}{Gbit/s}. The remaining
incoming link serves as a spare clock input, three of the outgoing links provide spare
bandwidth for upgrades, as does a second firefly socket that is left unequipped per default.

The Arria~V is configured and controlled from an Intel MAX~10 FPGA (capable of configuring
itself from built in flash memory). On power-up the MAX~10 reads the Arria configuration data
from an external quad-SPI flash memory. The flash memory can be written by the MAX~10 using
data received from a JTAG interface (only available with open detector) or from the crate controller
and the backplane (slow) or from the configuration optical link, via the Arria and an inter-FPGA SPI 
interface (fast, but requiring programmed Arria). The MAX~10 also offers an ADC with a multiplexer, 
which is used to monitor the supply voltages on the board and to read several strategically placed 
analogue temperature sensors.

The front-end board is powered with \SI{20}{V} from the backplane. Three switching DC/DC 
converters produce the \SI{3.3}{V}, \SI{2.5}{V} and \SI{1.1}{V} needed on the board; the
other voltages required are generated using linear regulators. As the boards are operated
inside a \SI{1}{T} solenoid, the inductors for the switching converters cannot have ferrite
cores; we employ air-core solenoids inside copper shielding boxes 
\cite{hesping2019, gagneur2020}. The overall power consumption is below \SI{10}{W}.
The boards are cooled with custom aluminium plates in thermal contact with the main
heat-producing components. A heat-pipe connects the aluminium plate with the water-cooled
SSW crate.

\subsection{Switching board}

\begin{figure}
  \centering
  \includegraphics[width=\columnwidth]{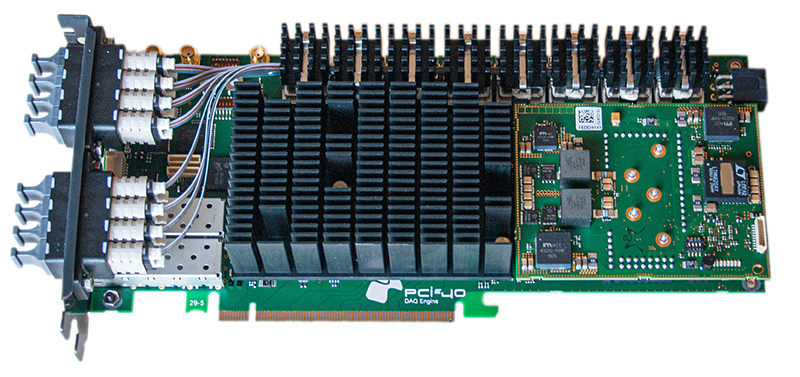}
  \caption{PCIe40 switching board. Optical fibres are routed from the MTP connectors
  on the left to eight Avago MiniPod optical receivers/transmitters. The Intel Arria~10
  FPGA as well as two MAX~10 configuration devices and the PCIe switch are covered by 
  the heat sink, the mezzanine card on the right provides the power.}
  \label{fig:switchingboard}
  \end{figure}

After about \SI{50}{m} of multi-mode optical fibre, the optical cables from the
front-end boards reach the Mu3e counting house, where a large fibre patch panel 
separates the clock and reset fibres from the control and data fibres. The latter
are routed to four PCIe40 \emph{switching boards}; see \figurename~\ref{fig:switchingboard}. 
These cards, developed in Marseille for the LHCb and ALICE upgrades \cite{cachemiche:2016reb}, 
provide 48 high-speed optical inputs and outputs using a total of eight Avago MiniPods. 
An Intel Arria~10 FPGA provides two generation 3, 8-lane PCIe interfaces.

On the switching board FPGA, the data streams from up to 34 front-end boards are time-aligned
and merged. For the fibre detector, coincidences between the signals from the two fibre
ends are formed to further suppress SiPM dark counts.
The resulting data streams for all detectors are then forwarded to the filter farm 
using \SI{10}{Gbit/s} optical links. 
Matching the bandwidth requirements (see Table~\ref{tab:FPGABandwidthRequirements}), the
central pixel detector uses eight links, the fibre detector uses four links and the forward and backward
combined pixel and tile detector stations use two links each.
We use custom protocols on all fibre links tailored to make maximum use of the bandwidth
given that we have just three hit types, all with well-defined sizes and formats.

For each data link from the front-end boards, there is a \SI{6.25}{Gbit/s} control link going in the
opposite direction. This is used for configuration data (such as the threshold tune values for more
than 180 million pixels) and to request monitoring information such as temperature values from the
front-end boards. The switching board FPGA is also used to extract information such as hit maps and
other histograms from the incoming data streams, which is made available to the MIDAS DAQ via
the PCIe interfaces.

All the Arria 10 FPGAs in the experiment feature generation 3, 8-lane PCIe interfaces with
common firmware, Linux kernel driver and software. The PCIe firmware provides four 
\emph{Bus-Addressable Register} (BAR) areas. The first BAR provides 64 \SI{32}{Bit} registers
writeable from the FPGA, the second BAR 64 registers writeable from the PC, the third and
fourth BARs are \SI{256}{Kbyte} memories, one writeable from the FPGA, one from the PC.
This is complemented by a \emph{direct memory access} (DMA) engine for fast data transfers
to the PC main memory. The DMA buffer on the PC is treated as a large ring buffer.
Signalling to the PC occurs without interrupts by performing DMA to a separate 
\emph{control memory} area containing pointers to the last written blocks \cite{dissvombruch2017}.
In tests of our DMA firmware, we can sustain \SI{38}{Gbit/s} of user data transfers 
\cite{koeppel2019}.

\subsection{Filter farm}

\begin{figure}
  \centering
  \includegraphics[width=\columnwidth]{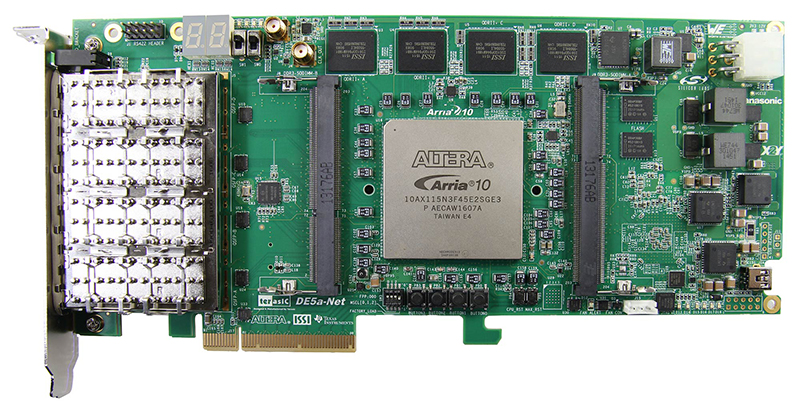}
  \caption{DE5aNET receiving board. The four QSFP quad optical transceivers are located on the
  left. In the centre is the Intel Arria~10 FPGA, flanked left and right by the DDR memory connectors
  and flash memory on the top. The PCIe edge connector is in the lower left and the DC/DC converters on the
  right. The MAX~10 configuration FPGA is on the back of the board.}
  \label{fig:receivingboard}
  \end{figure}

  \begin{figure}
    \centering
    \includegraphics[width=\columnwidth]{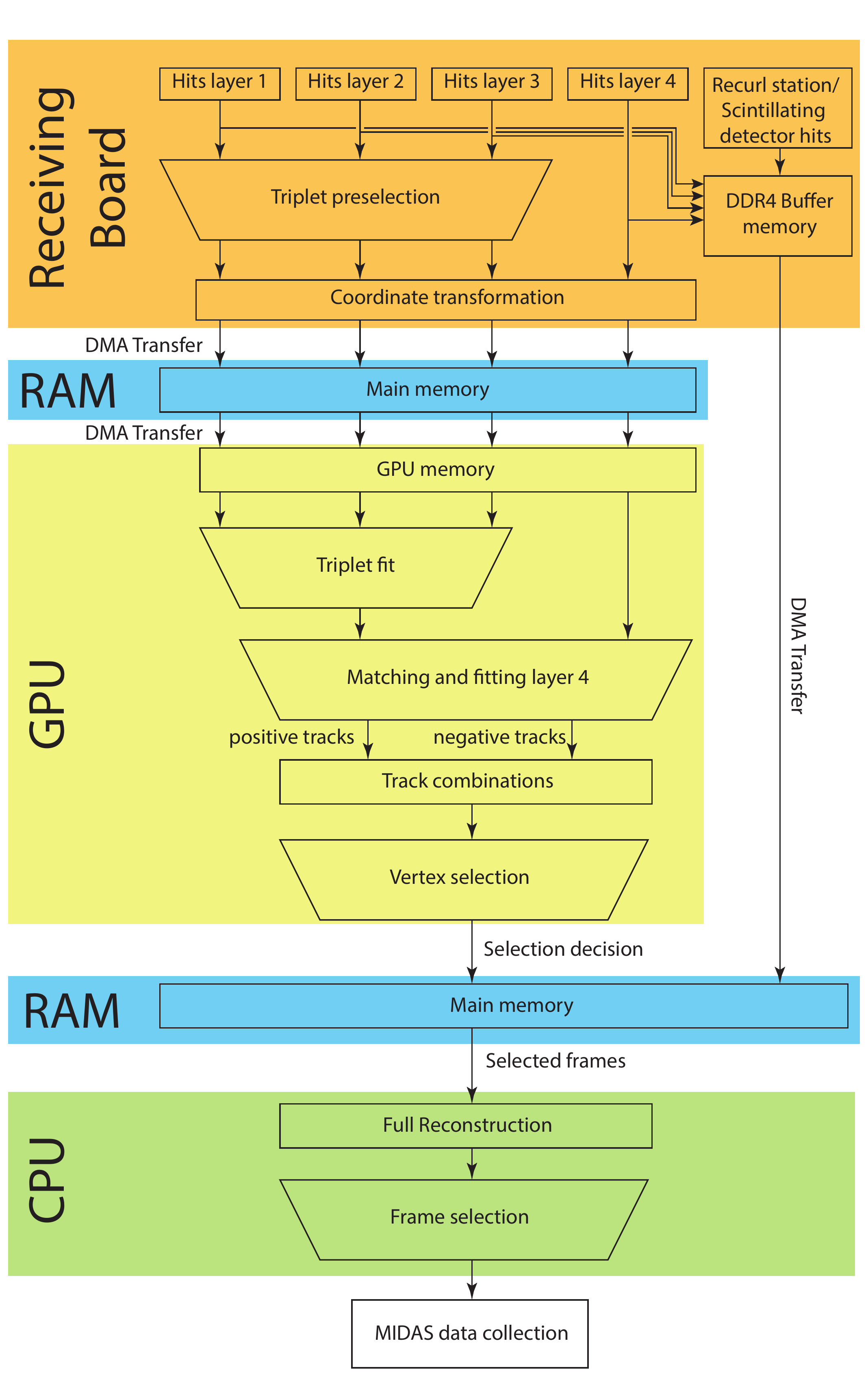}
    \caption{Data flow in the online reconstruction. Trapezoids are used to indicate selection decisions.}
    \label{fig:dataflow}
    \end{figure}

The sixteen \SI{10}{Gbit/s} links from the switching boards are routed to the Mu3e filter farm,
which consists of twelve PCs in a daisy chain configuration. Each PC is equipped with a 
\emph{receiving board}. Here we use the commercial DE5aNET board \cite{de5anet} produced 
by Terasic Inc. It features an Intel Arria~10 FPGA connected to four QSFP quad-optical transceivers
and two banks of DDR4\footnote{Earlier versions feature DDR3 memory.} memory. It connects to the host PC
using one 8-lane generation 3 PCIe interface.

\figurename~\ref{fig:dataflow} shows a schematic of the data flow in a single farm node.
The data from the sixteen links are received, time aligned and buffered to the DDR4 memory.  
If the buffers are full, the data are forwarded to the next PC in the daisy chain, providing a simple
form of load distribution without the need for a back-channel and back-pressure. 
The hits of the central pixel detector are extracted from the data stream,
and a first geometrical selection is performed based on hits from the first three detector layers;
only hit triplets compatible with stemming from a track originating in the target region are 
processed further. 
Using look-up memories, the hit coordinates are transformed
from the \SI{32}{Bit} sensor/column/row scheme to the global detector coordinate system (using three 
single-precision floating point numbers). The hits are then sent to the main memory of the host
PC using DMA transfers. From the main memory, the hit data are then forwarded to a
GPU\footnote{Due to the rapid developments of the GPU market at the time of writing in terms of
both prices and performance, we will choose the exact model of GPU for production use as late as 
possible. Currently we aim for mass-market gaming cards.}. On the GPU, a custom developed track fit
treating multiple scattering as the only uncertainty \cite{berger:2016vak} is performed on hits
from the first three detector layers. If successful, the track is extrapolated to the fourth
layer and if a matching hit is found, it is included in the fit, or otherwise the track is discarded.
Lists of tracks with positive and negative charge assignments are created and a simple vertex
fit based on circle intersections is performed for all combinations
of two positive and one negative track (corresponding to the signal topology) \cite{dissvombruch2017}. If the three tracks are
compatible with originating in a common vertex, the frame is tagged for readout. 
In this case the complete detector information for the frame is 
retrieved from the DDR4 buffer, passed to the main memory of the host PC and a full reconstruction
\cite{kozlinskiy:2017wyl} is performed. Signal candidate events are saved using the MIDAS event
handling infrastructure to the \emph{PetaByte Archive} operated jointly by PSI and the Swiss
supercomputing centre CSCS. We aim for an output data rate of about \SI{100}{MB/s} to keep 
storage requirements manageable and affordable. If the selection criteria are relaxed, the
DAQ system is easily capable of saturating the bandwidth of both local storage media or the
outgoing Ethernet connection.

As the muon decays on the target have no particular time structure, the online reconstruction on
the GPUs is performed in overlapping time frames\footnote{The size of the overlap will be determined 
once the exact time resolution of the production pixel sensor is known.} of \SI{64}{ns} length in order to avoid an efficiency
loss at the frame boundary. The frame size is chosen to cover at least $3\sigma$ of the pixel time 
resolution and has to be a multiple of the time stamp bin size.

The Mu3e detector is peculiar in as much as the reconstructed information (i.e.~the helix parameters
of a track represented as floating-point numbers) takes more space than the raw data. This, together 
with the high rates, precludes saving e.g.~track kinematics for all events. What can however be done is 
the histogramming of such quantities on the GPUs. Given a good calibration of the online reconstruction, 
these histograms can be used in physics analyses, e.g.~searches for two-body decays of the muon 
$\mu \rightarrow e X$, where Mu3e has a very competitive sensitivity 
\cite{perrevoort:2018okj,Perrevoort:2018ttp, calibbi:2020jvd}.

\subsection{Synchronisation}

\begin{figure}
  \centering
  \includegraphics[width=\columnwidth]{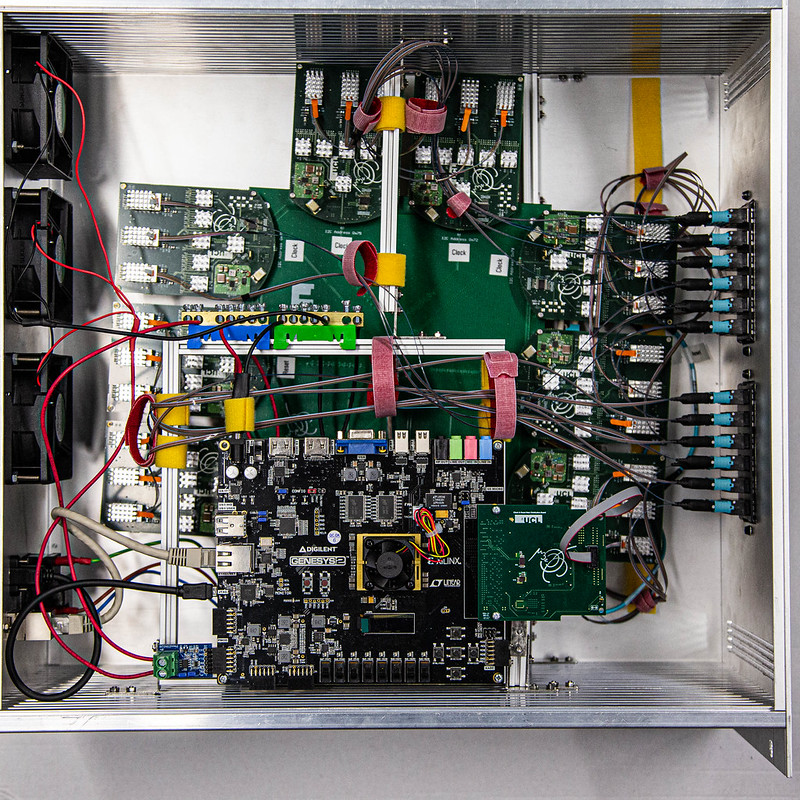}
  \caption{Clock and reset distribution box. Bottom centre is the Genesys-2 FPGA board,
  partly covering the distribution mother board with eight daughter boards attached (three on the
  left and right, two on the top). Power, control signals and forced air flow enter on the left,
  the optical clock and reset signals exit on the right.}
  \label{fig:clockbox}
  \end{figure}

The Mu3e detector and DAQ are all synchronised to a \SI{125}{MHz} master clock, and all other clocks,
e.g.~the \SI{625}{MHz} clock for the MuTRiG TDC ASICs or the optical link clocks are derived
from this master clock using \emph{Phase-Locked Loops} (PLLs). In order to also determine a common
starting point for time stamps, a common reset signal is distributed to all parts of the experiment.
For this we employ the \SI{1.25}{Gbit/s}, 8b/10b encoded \emph{reset stream}, which we can also 
use to signal state transitions such as run starts and stops.

The reset and clock are generated and fanned-out in a single four unit 19~inch box, the
\emph{clock and reset system}; see \figurename~\ref{fig:clockbox}.
The \SI{125}{MHz} master clock is generated by a Silicon Labs SI5345 clock generation IC. The 
reset stream is generated using a gigabit transceiver on a commercial Digilent Genesys~2 board 
\cite{genesys}. This board, featuring a 
Xilinx Kintex-7 FPGA, is also used to configure and monitor the clock and reset system.
The modular system uses one mother- and eight daughter-boards equipped with ON-Semiconductor 
NB7L1008M differential eight-fold fan-out chips. Samtec Firefly transmitters are used to generate
the 144 optical clock copies and the 144 optical reset stream copies.
A jitter of less than \SI{5}{ps} between the different output clocks (after optical 
transmission and back-conversion to differential electrical signals) was measured, easily fulfilling
the \SI{30}{ps} specification.

\section{Conclusion}
We have presented the design and hardware implementation of the Mu3e data acquisition,
a compact and cost effective system capable of dealing with rates in excess of
\SI{100}{Gbit/s}. The fully streaming system employs a custom front-end board inside the
detector magnet collecting the detector ASIC data and forwarding them optically to 
switching boards in the counting house, which also interface to the detector control and
monitoring. A small filter farm employing FPGA receiving boards and consumer GPUs performs
a full track reconstruction and selects events of interest. All components of the Mu3e
DAQ are available and were successfully tested. The full system is expected to be set up
a PSI and operational by the end of 2021.


%

\ifCLASSOPTIONcaptionsoff
  \newpage
\fi

\bibliographystyle{IEEEtran}

\bibliography{DAQIEEE}

\end{document}